\documentclass{article}
\usepackage{amsfonts}
\usepackage{amssymb}
\usepackage{amsmath}

\input{tcilatex}
\begin{document}

\title{Entropic Dynamics and Quantum \textquotedblleft
Measurement\textquotedblright \thanks{%
Invited paper presented at the 41st International Conference on Bayesian and
Maximum Entropy Methods in Science and Engineering -- MaxEnt2022 (July
18-22, Institut Henri Poincar\'{e}, Paris)}}
\author{Ariel Caticha \\
{\small Physics Department, University at Albany-SUNY, Albany, NY 12222, USA.%
}}
\date{ }
\maketitle

\begin{abstract}
The entropic dynamics (ED) approach to quantum mechanics is ideally suited
to address the problem of measurement because it is based on entropic and
Bayesian methods of inference that have been designed to process information
and data. The approach succeeds because ED achieves a clear-cut separation
between ontic and epistemic elements: positions are ontic while
probabilities and wave functions are epistemic. Thus, ED is a viable\emph{\
realist }$\psi $\emph{-epistemic model}. Such models are widely assumed to
be ruled out by various no-go theorems. We show that ED evades those
theorems by adopting a purely epistemic dynamics and denying the existence
of an ontic dynamics at the subquantum level.
\end{abstract}

\section{Introduction}

A measurement is a physical process like any other and, therefore, its
analysis should cause no difficulties once a proper understanding of the
relevant dynamics has been achieved \cite{Bell 1990b}. Nevertheless, the
problem of quantum measurement has historically been a source of endless
controversy. It is intimately associated with most of those features of
quantum mechanics (QM) that make it so strange and fascinating (see \emph{%
e.g.}, \cite{Ballentine 1998}-\cite{Leifer 2014}). Does the quantum state
reflect incomplete information or is it something real, ontic? If the
latter, can wave functions undergo a physical collapse during measurement?
Alternatively, if no collapses ever occur, and wave functions always obey
the linear Schr\"{o}dinger equation, how could quantum measurements ever
yield definite outcomes? How does one negotiate the interface between the
microscopic quantum world and the macroscopic classical world of the
measuring device? Do at least some privileged variables represent something
real with definite values at all times? Or, alternatively, are the values of
all observables created during the act of measurement? If so, how can one
ever say that anything real exists when nobody is looking?

Our first goal here is to address the problem of measurement from the
perspective of Entropic Dynamics (ED) \cite{Caticha 2019}\cite{Caticha 2021}%
. The ED approach to QM is ideally suited to tackle the questions above
because it is based on entropic and Bayesian methods of inference that have
been designed to process information and data. (For a more detailed
presentation see \cite{Caticha 2022}.) The success of the ED approach hinges
on a clear ontological and epistemic commitment: In ED the positions of
particles enjoy the privileged role of being the only ontic variables.
Indeed, all measurements are ultimately position measurements --- whether
they consist of a direct measurement of a particle's position or an indirect
measurement of the position of a pointer variable. In contrast,
probabilities and wave functions are fully epistemic in nature. Thus,
positions reflect real properties with definite values that are not created
by the act of measurement while all other observables are epistemic because
they reflect properties of the wave function \cite{Johnson Caticha 2011}\cite%
{Vanslette Caticha 2016}. This explains how it is that their values are
\textquotedblleft created\textquotedblright\ by the act of measurement.

Models such as ED that invoke ontic variables while the wave function
remains an epistemic object are described as \textquotedblleft realist $\psi 
$-epistemic models\textquotedblright . In contrast, the various descendants
of Bohr's Copenhagen interpretation which deny a definite quantum reality
are dubbed \textquotedblleft anti-realist $\psi $-epistemic
models\textquotedblright , while models such as the de Broglie-Bohm and
many-worlds models are called \textquotedblleft realist $\psi $-ontic
models\textquotedblright .

There exist a number of powerful no-go theorems --- the so-called $\psi $%
-ontology theorems \cite{Leifer 2014} --- that rule out large families of $%
\psi $-epistemic \textquotedblleft ontological\textquotedblright\ models \ 
\cite{Bell 1966}{}\cite{Spekkens 2005}\cite{Harrigan Spekkens 2010} because
they disagree with QM (\emph{e.g. }\cite{Bell 1966}\cite{Bell 1990a}-\cite%
{Tumulka 2022}. (The term `ontological models' has been proposed as an
improved way to refer to the old `hidden-variable models'. The new term
recognizes that some \textquotedblleft hidden\textquotedblright\ variables,
such as positions, are observable and, therefore, not at all hidden.) These
no-go theorems have been interpreted as strong evidence in favor of
\textquotedblleft realist $\psi $-ontic models\textquotedblright . However, $%
\psi $-epistemic models remain highly appealing, not least because they
trivially explain the infamous wave function collapse as a mere updating of
probabilities in the light of new data. Remarkably, fully developed realist $%
\psi $-epistemic models\ such as ED are scarce \cite{Leifer 2014}. To my
knowledge ED is the only such model that provides a detailed reconstruction
of the formalism of QM and claims to reproduce not just a fragment of
quantum phenomena, but QM in its totality.

A second goal of this paper is to analyze how ED evades the consequences of $%
\psi $-ontology theorems. Ontological models assume the existence of ontic
variables. We shall argue that they also implicitly assume the existence of
some ontic dynamics at the subquantum level. (The details of the dynamics
need not be specified and therein lies the power and generality of the $\psi 
$-ontology theorems.) ED, on the other hand, makes a commitment to ontic
variables while denying them an ontic dynamics; ED is a purely epistemic
dynamics of probabilities. There is no implication that particles move as
they do because they are pushed around by other particles or guided by an
ontic pilot wave. What wave functions do is to guide our expectations about
where particles might be found but there is no mechanism that accounts for
any causal influence on the particles themselves. ED is a \emph{mechanics
without a mechanism}.

Section 2 contains a brief overview of ED. In section 3 I discuss the direct
measurement of microscopic positions including their amplification to
achieve observability at the macroscopic level \cite{Johnson Caticha 2011}.
Then, in section 4, I discuss how the use of more elaborate devices allows
us to define other non-position observables. I show how their
\textquotedblleft measurement\textquotedblright\ is ultimately reduced to
the direct measurement of positions and derive the associated Born rule \cite%
{Johnson Caticha 2011}\cite{Caticha 2000}. The special case of the classical
von Neumann measurements provides an interesting extension \cite{Vanslette
Caticha 2016}. Thus far these sections review our previous work on this
subject. In section 5 I present new material that addresses the question of
how ED manages to evade the various $\psi $-ontology theorems.

\section{Brief review of entropic dynamics}

To set the context for the rest of the paper we review the main ideas that
form the foundation of entropic dynamics. For a detailed account see \cite%
{Caticha 2019}-\cite{Caticha 2022}. Here is a brief summary:

\noindent \textbf{Ontological clarity:} Particles have definite but unknown
positions $\{x^{A}\}$ collectively denoted by\emph{\ }$x$. These are the
ontic microstates. ($A$ is a composite index, $A=(n,a)$, where $n$ $=1\ldots
N$ labels the particles, and $a=1,2,3$ the three spatial coordinates.) The
particles follow continuous trajectories and the goal is to predict the
probability $\rho (x)$ of the positions $x$ on the basis of some limited
information.

\noindent \textbf{ED is a dynamics of probabilities:} The probability of a
step from $x$ to a neighboring $x^{\prime }$, $P(x^{\prime }|x)$, is found
by maximizing its entropy relative to a prior that enforces short steps and
subject to appropriate constraints that introduce directionality and
correlations. The main constraint involves a function $\phi (x)$ that plays
three separate roles: first, it is related to a constraint in the
maximization of entropy; second, if the probabilities $\rho (x)$ are
considered as generalized coordinates, then $\phi (x)$ is the momentum that
is canonically conjugate to them; and third, $\phi (x)$ is the phase of the
quantum wave function, $\psi =\rho ^{1/2}e^{i\phi /\hbar }$.

\noindent \textbf{Entropic time:} An epistemic dynamics of probabilities
inevitably leads to an epistemic notion of time. The construction of time
involves the introduction of the concept of an instant, the notion that the
instants are suitably ordered, and a convenient definition of duration. By
its very construction there is a natural arrow of entropic time.

\noindent \textbf{The evolution of probabilities} is found by the
accumulation of the short steps described by $P(x^{\prime }|x)$. This
results in a continuity equation that is local in configuration space but
leads to non-local correlations in physical space,%
\begin{equation}
\partial _{t}\rho _{t}(x)=-\partial _{A}\left( \rho _{t}v^{A}\right) \quad 
\text{where}\quad v^{A}=m^{AB}\partial _{B}\phi ~.  \label{cont eq}
\end{equation}%
Notation: $\partial _{A}=\partial /\partial x^{A}$; $m_{AB}=m_{n}\delta
_{AB} $ is the mass tensor, $m_{n}$ are the particle masses, and $%
m^{AB}=\delta ^{AB}/m_{n}$ is the inverse mass tensor.)

\noindent \textbf{Symplectic structure:} For a suitable choice of a
functional $\tilde{H}[\rho ,\phi ]$ the continuity equation (\ref{cont eq})
can be written in Hamiltonian form, 
\begin{equation}
\partial _{t}\rho _{t}(x)=\frac{\delta \tilde{H}}{\delta \phi (x)}~,
\end{equation}%
which suggests choosing $(\rho ,\phi )$ as a pair of canonically conjugate
variables. The epistemic phase space (or e-phase space) $\{\rho ,\phi \}$
has a natural symplectic structure with symplectic two-form $\Omega $.

\noindent \textbf{Information geometry:}\emph{\ }The e-phase space is
assigned a metric structure with metric tensor $G$ based on the information
metric of the statistical manifold $\{\rho \}$ of probabilities $\rho (x)$.
The joint presence of symplectic and metric structures implies the existence
of a complex structure and suggests the introduction of wave functions $\psi
=\rho ^{1/2}e^{i\phi /\hbar }$ as complex coordinates. (For a discussion of
the subtleties concerning the correct choice of the spaces that are
cotangent to the manifold $\{\rho \}$ and of the metric structure associated
to e-phase space $\{\rho ,\phi \}$ see \cite{Caticha 2021}.)

\noindent \textbf{The epistemic dynamics} that preserves the symplectic
structure in the sense of vanishing Lie derivative, $\pounds _{H}\Omega =0$,
obeys Hamilton's equations, 
\begin{equation}
\partial _{t}\rho (x)=\frac{\delta \tilde{H}}{\delta \phi (x)}~,\quad
\partial _{t}\phi (x)=-\frac{\delta \tilde{H}}{\delta \rho (x)}~.~
\label{Hamilton eqs}
\end{equation}%
If we further require the preservation of the metric structure, $\pounds %
_{H}G=0$, and of the normalization of probabilities we find that $\tilde{H}$
is constrained to be bilinear in $\psi $ and $\psi ^{\ast }$, 
\begin{equation}
\tilde{H}[\psi ,\psi ^{\ast }]=\int d^{3N}xd^{3N}x^{\prime }\,\psi ^{\ast
}(x)\hat{H}(x,x^{\prime })\psi (x^{\prime })~,  \label{bilinear hamiltonian}
\end{equation}%
which implies that (\ref{Hamilton eqs}) can be rewritten as a \emph{linear}
Schr\"{o}dinger equation,%
\begin{equation}
i\hbar \frac{d\psi (x)}{dt}=\int d^{3N}x^{\prime }\,\hat{H}(x,x^{\prime
})\psi (x^{\prime })~.  \label{sch a}
\end{equation}%
The particular form of the Hamiltonian kernel $\hat{H}(x,x^{\prime })$ is
determined by requiring that it reproduce the ED evolution of probabilities,
eq.(\ref{cont eq}). In standard notation we find 
\begin{equation}
i\hbar \partial _{t}\psi =\tsum\nolimits_{n}\frac{-\hbar ^{2}}{2m_{n}}\nabla
_{n}^{2}\psi +V(x)\psi ~.  \label{sch b}
\end{equation}%
\textbf{Entropic dynamics} is the purely epistemic dynamics of $(\rho ,\phi
) $ or, equivalently, of $\psi $; there is no underlying ontic dynamics of $%
x $. Compared to other models of QM ED is fairly conservative in that it
confers ontic status to configurational variables such as position and a
clear epistemic status to probabilities, phases, and wave functions. But ED
is radically non-classical in that it denies the ontic status of dynamics
and of all observables (energy, momentum, etc.) except position.

\noindent \textbf{Hilbert space:} To conclude the reconstruction of QM we
can take full advantage of the linearity of (\ref{sch b}) and introduce
Hilbert spaces and the Dirac notation: 
\begin{equation}
|\psi \rangle =\int d^{3N}x\,|x\rangle \psi (x)~,\quad \psi (x)=\langle
x|\psi \rangle ~,\quad \text{and}\quad \hat{H}(x,x^{\prime })=\langle x|\hat{%
H}|x^{\prime }\rangle ~.
\end{equation}

\section{Measuring position: amplification}

All measurements are position measurements. The measurement of the position
of a microscopic particle is conceptually straightforward because \emph{the
particle already has a definite position }$x$\emph{.} The issue of inferring 
$x$ is not different from the way data information is handled in any other
Bayesian inference problem. There is however the technical problem of
amplifying microscopic details so they can become macroscopically
observable. This is usually handled with a detection device set up in an
initial unstable equilibrium. For example, QM allows us to calculate the
probability $P(a|x)$ that a particle at $x$ will ionize a neighboring atom
located at $a$. In a bubble chamber the ionized atom will trigger the
formation of a bubble centered at $a$; in a photographic emulsion the ion
will trigger the formation of a silver crystallite centered at $a$. More
generally, the particle activates the amplifying system by inducing a
cascade reaction that leaves the amplifier in a definite macroscopic final
state described by some \textquotedblleft pointer\textquotedblright\
variable $a$.

The goal of the amplification process is to allow us to infer the
microscopic position $x$ from the observed macroscopic position $a$ of the
pointer variable. Incidentally, the latter is just a classical variable \cite%
{Demme Caticha 2016}. Once the likelihood function $P(a|x)$ is given, the
value $x$ can be inferred following a standard application of Bayes rule, 
\begin{equation}
P(x|a)=P(x)\frac{P(a|x)}{P(a)}\ .
\end{equation}%
In practice life is more complicated and the likelihood function will be
distorted and smeared by spurious correlations and noise. A successful
measurement always involves, of course, a skilled experimentalist who will
design the device so that those unwanted effects will be minimized and
controlled.

The point of these considerations is to emphasize that there is nothing
intrinsically quantum mechanical about the amplification process.

\section{Defining and \textquotedblleft measuring\textquotedblright\ other
observables}

Position is easy because it is an ontic quantity. Next we tackle observables
other than position: how they are defined and how they are measured \cite%
{Caticha 2000}\cite{Johnson Caticha 2011}. For notational convenience we
initially consider the case of a single particle that lives on a lattice;
the measurement of its position leads to a discrete set $x_{k}$ of possible
outcomes. The translation from continuous to discrete positions is
straightforward, 
\begin{equation}
\psi (x)=\rho ^{1/2}(x)e^{i\phi (x)/\hbar }\quad \text{becomes}\quad \psi
_{k}=p_{k}^{1/2}e^{i\phi _{k}/\hbar }\ ,
\end{equation}%
and%
\begin{equation}
\rho (x)\,d^{3}x=|\langle x|\psi \rangle |^{2}\,d^{3}x\quad \text{becomes}%
\quad p_{k}=|\langle x_{k}|\psi \rangle |^{2}\ .
\end{equation}%
If the state is 
\begin{equation}
|\psi \rangle =\sum\nolimits_{k}c_{k}|x_{k}\rangle \quad \text{then}\quad
p_{k}=|\langle x_{k}|\psi \rangle |^{2}=|c_{k}|^{2}~.  \label{Born a}
\end{equation}%
Since position is the only ontic quantity it is not strictly necessary to
define other observables except that they turn out to be convenient to
discuss more complex experiments in which the particle is subjected to
additional interactions, such as magnetic fields or diffraction gratings,
before it reaches the position detectors.

The fact that measurements are dynamical processes means that the
interactions within a complex measurement device $\mathcal{M}$ are described
by a linear and unitary evolution $\hat{U}_{M}$ given by a Hamiltonian $\hat{%
H}_{M}$. The particle will be detected with certainty at position $%
|x_{k}\rangle $ provided it was initially in a state $|s_{k}\rangle $ such
that 
\begin{equation}
\hat{U}_{M}|s_{k}\rangle =|x_{k}\rangle \ .  \label{unitary evolution}
\end{equation}%
Since the set $\{|x_{k}\rangle \}$ is orthonormal and complete, the
corresponding set $\{|s_{k}\rangle \}$ is also orthonormal and complete, 
\begin{equation}
\langle s_{j}|s_{k}\rangle =\delta _{jk}\quad \text{and}\quad
\sum\nolimits_{k}|s_{k}\rangle \langle s_{k}|{}=\hat{I}\ .
\label{completeness}
\end{equation}%
To find the effect of the complex device $\mathcal{M}$ on a generic initial
state $|\psi \rangle $ we express it in the $\{|s_{k}\rangle \}$ basis, 
\begin{equation}
|\psi \rangle =\sum\nolimits_{k}c_{k}|s_{k}\rangle \ ,
\end{equation}%
where the complex coefficients $c_{k}=\langle s_{k}|\psi \rangle $ are
normalized. The state $|\psi \rangle $ evolves through $\mathcal{M}$
according to $\hat{U}_{M}$ so that when it reaches the actual position
detectors the new state is 
\begin{equation}
\hat{U}_{M}|\psi \rangle =\sum\nolimits_{k}c_{k}\hat{U}_{M}|s_{k}\rangle
=\sum\nolimits_{k}c_{k}|x_{k}\rangle \ .
\end{equation}%
According to the Born rule for position measurements, eq.(\ref{Born a}), the
probability of finding the particle at the position $x_{k}$ is 
\begin{equation}
p_{k}=|c_{k}|^{2}=|\langle s_{k}|\psi \rangle |^{2}\ .  \label{Born b}
\end{equation}%
In words: The probability that the particle in the initial epistemic state $%
|\psi \rangle $ is later found at position $x_{k}$ is $|c_{k}|^{2}$.

Note that \emph{the particle in the initial epistemic state }$|\psi \rangle $%
\emph{\ has been detected in ontic state }$x_{k}$\emph{\ as if it had
earlier been in the epistemic state }$|s_{k}\rangle $\emph{.} The argument
above illustrates the main idea. Generalizations such as, for example, to
continuous configuration spaces, are straightforward \cite{Caticha 2000}. In
this case the (suitably amplified) position $x_{k}$ plays the role of a
pointer variable but has anything been \textquotedblleft
measured\textquotedblright\ here? \cite{Bell 1990b}.

This process can be described in a slightly different language that
unfortunately obscures the distinction between the ontic nature of $x_{k}$
and the epistemic nature of $|x_{k}\rangle $ (or, more generally, of $\psi
_{k}=\langle x_{k}|\psi \rangle $). We shall say that \emph{the particle is
detected in state }$|x_{k}\rangle $\emph{\ as if it had earlier been in the
state }$|s_{k}\rangle $\emph{. }We can further obscure the language by
de-emphasizing the inner workings of the complex device, forgetting the
dynamics, and treating the detector as a black box. The result is a more
concise and more misleading statement: \emph{the particle has been
\textquotedblleft detected\textquotedblright\ in the state }$|s_{k}\rangle $%
. Continuing along the same lines leads us to adopt the language that is
standard in QM textbooks: \emph{the probability that the particle in state }$%
|\psi \rangle $\emph{\ is \textquotedblleft detected\textquotedblright\ in
state }$|s_{k}\rangle $\emph{\ is }$|\langle s_{k}|\psi \rangle |^{2}$ ---
which reproduces Born's rule for a generic measurement device $\mathcal{M}$.
But, by now, the real meaning of what has been `detected' lies buried deep
underground.

The same complex detector $\mathcal{M}$ can be used to \textquotedblleft
measure\textquotedblright\ all operators of the form 
\begin{equation}
\hat{M}=\sum\nolimits_{k}\alpha _{k}|s_{k}\rangle \langle s_{k}|
\label{operator M}
\end{equation}
where the eigenvalues $\alpha _{k}$ are arbitrary scalars. This establishes
the eigenvector-eigenvalue connection. Note that when we say we have
detected the particle at $x_{k}$ \emph{as if} it had earlier been in state $%
|s_{k}\rangle $ we are absolutely not implying that the particle \emph{was}
in the particular epistemic state $|s_{k}\rangle $ --- this is just a figure
of speech. The actual epistemic state was $|\psi \rangle $ not $%
|s_{k}\rangle $. When the system is \textquotedblleft detected in $%
|s_{k}\rangle $\textquotedblright\ the standard language is that the outcome
of the measurement is the eigenvalue $\alpha _{k}$. It is then clear that
the outcome $\alpha _{k}$ was not a pre-existing value and it is in this
sense that one says that the value $\alpha _{k}$ was \textquotedblleft
created by the act of measurement\textquotedblright .

This point deserves to be made more explicit: sentences such as
\textquotedblleft the particle has momentum $\vec{p}\,$\textquotedblright\
or \textquotedblleft it has energy $E$\textquotedblright\ are to be
recognized as mere linguistic shortcuts that convey information about
components of the wave function before the particle enters the complex
detector. Therefore, strictly speaking, there is no such thing as the
momentum or the energy \emph{of the particle}: the momentum and the energy
are not properties of the particle but properties of special epistemic
states.

In the standard language one refers to the operator $\hat{M}$ as
representing an \textquotedblleft observable\textquotedblright\ and it is
common to attribute to its eigenvalues and eigenvectors the status of being
ontic --- actual physical properties. This is not a mere abuse of language;
in ED it is just plain wrong. Since what one is actually doing is inferring
properties of the wave function from measurements of position, a more
appropriate terminology is to refer to $\hat{M}$ as an \textquotedblleft
inferable\textquotedblright\ \cite{Vanslette Caticha 2016}.

To summarize: In the standard interpretation of quantum mechanics Born's
rule for generic measurements is a postulate. In ED it is the natural
consequence of a unitary time evolution and the hypothesis that \emph{all
measurements are ultimately position measurements}.

\subsubsection*{An illustration: von Neumann measurements}

So far we have just discussed measurements that rely on the direct detection
of the position of the particle (and its subsequent amplification). One can
substantially enlarge the class of useful experiments by considering complex
setups in which one infers properties of one system indirectly by measuring
the position of another system --- the pointer variable --- with which the
system has interacted. Nothing in this section is original material; it is
included merely as a purely pedagogical illustration of the fact that all
measurements are position measurements.

The system of interest is composed of one or many particles; its ontic state
is $x=\{x_{n}\}$ and its epistemic state is $|\psi \rangle $. The pointer
device is also a particle; its ontic state is $X$ and its epistemic state is 
$|\pi \rangle $. The interaction between the system and the pointer is
described by a Hamiltonian modelled as 
\begin{equation}
\hat{H}_{M}=-g(t)\hat{P}\hat{M}~,  \label{H a}
\end{equation}%
where $\hat{M}=$ $\sum\nolimits_{k}\alpha _{k}|s_{k}\rangle \langle s_{k}|$
is the operator to be \textquotedblleft measured,\textquotedblright\ and $%
\hat{P}$ is the operator that generates translations of the pointer states,%
\begin{equation}
e^{-i\hat{P}\alpha /\hbar }|X\rangle =|X+\alpha \rangle ~.
\end{equation}%
The function $g(t)$ measures the strength of the interaction. We make the
usual assumptions: (a) that $\tint g(t)dt=1$, (b) that $g(t)$ vanishes
before and after the measurement, and (c) that while the measurement lasts $%
g(t)$ it is large enough that $\hat{H}_{M}$ is a good approximation to the
full Hamiltonian.

The pointer is set to its initial \textquotedblleft ready\textquotedblright\
position near $X_{i}=0$ with some uncertainty $\sigma _{\pi }$, 
\begin{equation}
|\pi \rangle =N_{X}\int \,dX_{i}\,e^{-X_{i}^{2}/4\sigma _{\pi
}^{2}}|X\rangle ~,
\end{equation}%
where $N_{X}=(2\pi \sigma _{\pi }^{2})^{-3/4}$ is a normalization constant.
The initial state of the system is $|\psi \rangle
=\sum\nolimits_{k}c_{k}|s_{k}\rangle $. As a result of the interaction, the
system plus pointer evolve according to 
\begin{equation}
U_{M}|\psi \pi \rangle =\exp \left( -\frac{i}{\hbar }\int \hat{H}%
_{M}dt\right) |\psi \rangle |\pi \rangle =\exp \left( \frac{i}{\hbar }\hat{P}%
\hat{M}\right) |\psi \rangle |\pi \rangle ~,  \label{H c}
\end{equation}%
and become entangled. Using (\ref{H a})-(\ref{H c}) we find 
\begin{equation}
U_{M}|\psi \pi \rangle =N_{X}\sum\nolimits_{k}c_{k}\int
\,dX_{f}\,e^{-(X_{f}-\alpha _{k})^{2}/4\sigma _{\pi }^{2}}|s_{k}\rangle
|X_{f}\rangle ~,  \label{Born c}
\end{equation}%
which shows that the probability of the pointer position $X$ has been
shifted from an initial Gaussian centered at $X_{i}\approx 0$ to a final
mixture of Gaussians centered at $X_{f}\approx \alpha _{k}$, 
\begin{equation}
\Pr (X_{f})=\sum\nolimits_{k}|c_{k}|^{2}\frac{1}{(2\pi \sigma _{\pi
}^{2})^{3/2}}\,e^{-(X_{f}-\alpha _{k})^{2}/2\sigma _{\pi }^{2}}~.
\label{Born d}
\end{equation}%
When $\sigma _{\pi }$ is small and the Gaussian distributions are neatly
resolved we have a \textquotedblleft strong\textquotedblright\ or
\textquotedblleft von Neumann\textquotedblright\ measurement. The conclusion
is that measuring the final pointer position $X_{f}$ allows us to infer the
eigenvalue $\alpha _{k}$. (See the appendix for more details.)

When the Gaussian distributions overlap significantly the measurement is
said to be \textquotedblleft weak\textquotedblright . Such weak measurements
can nevertheless still be useful, not because they allow us to infer the
eigenvalues $\alpha _{k}$, but because they allow us to infer other
quantities such as the phase or even the wave function itself (for more on
this see \cite{Vanslette Caticha 2016} and references therein).

\section{Evading the no-go theorems}

The no-go theorems that rule out large families of realistic $\psi $%
-epistemic models are formulated in a framework of \textquotedblleft
ontological models\textquotedblright\ that originates with Bell \cite{Bell
1966}. The idea is that prior to an actual measurement the system undergoes
some sort of preparation procedure $\mathcal{P}$ and the result of the
measurement $\mathcal{M}$ is to yield outcomes labeled $k$ (inferred from
either $x$ or $X$ in the previous section). The goal is to find the
probability $p(k|\mathcal{M},\mathcal{P})$ of an outcome $k$ for given $%
\mathcal{P}$ and $\mathcal{M}$.

In realist models the ontic state of the system is denoted by variables $%
\lambda $. In ED, for example, $\lambda $ are the particle and pointer
positions $x$ and $X$, while in the de Broglie-Bohm model $\lambda $
consists of both $x$, $X$ and $\psi $. It is assumed that the preparation
procedure $\mathcal{P}$ may determine $\psi $ completely but it need not
yield complete control over $\lambda $ --- $\mathcal{P}$ only determines the
probability distribution, $p(\lambda |\mathcal{P})$. Thus, as the system
enters the measuring device $\mathcal{M}$, we are not only uncertain about
the future outcome $k$ but also of the initial values (\emph{i.e.}, before $%
\mathcal{M}$) of $\lambda _{i}$. This means that the relevant probability to
be discussed is the \emph{joint} distribution $p(k,\lambda _{i}|\mathcal{M},%
\mathcal{P})$ and the probability of the outcome $k$ is given by
marginalizing over $\lambda _{i}$,%
\begin{equation}
\tint d\lambda _{i}p(k,\lambda _{i}|\mathcal{M},\mathcal{P})=p(k|\mathcal{M},%
\mathcal{P})~.  \label{outcome prob a}
\end{equation}%
So far we have just used the rules of probability theory which, being of
universal applicability, also apply to QM. The desired goal is to find
realist models such that the distribution on the right of (\ref{outcome prob
a}) matches the predictions of QM such as eq.(\ref{Born b}).

To proceed further we write (\ref{outcome prob a}) as 
\begin{equation}
\tint d\lambda _{i}\,p(\lambda _{i}|\mathcal{M},\mathcal{P})p(k|\mathcal{M},%
\mathcal{P},\lambda _{i})=p(k|\mathcal{M},\mathcal{P})~
\label{outcome prob b}
\end{equation}%
and consider the two factors on the left separately. Concerning the first
factor, $p(\lambda |\mathcal{M},\mathcal{P})$, we shall assume that the
distribution of $\lambda _{i}$ is settled by the earlier choice of
preparation $\mathcal{P}$ and is independent of whatever choice one might 
\emph{later} make for the measurement device $\mathcal{M}$, 
\begin{equation}
p(\lambda _{i}|\mathcal{M},\mathcal{P})=p(\lambda _{i}|\mathcal{P})~.
\label{prep prob}
\end{equation}%
This is a statement of causality \cite{Bell 1990a}: conditional on $\mathcal{%
P}$, $\lambda _{i}$ is independent of the later choice of $\mathcal{M}$ .

The crucial assumption that defines what \cite{Spekkens 2005}\cite{Harrigan
Spekkens 2010} call an \textquotedblleft ontological
model\textquotedblright\ concerns the second factor, the response function $%
p(k|\mathcal{M},\mathcal{P},\lambda _{i})$. The assumption is that the
distribution of outcomes depends only on the ontic state $\lambda _{i}$ 
\emph{after} the preparation $\mathcal{P}$ but \emph{before} the device $%
\mathcal{M}$ and on the actual measurement performed, 
\begin{equation}
p(k|\mathcal{M},\mathcal{P},\lambda _{i})=p(k|\mathcal{M},\lambda _{i})~.
\label{outcome prob c}
\end{equation}%
This assumption does not necessarily violate QM. For example in the de
Broglie-Bohm model, $\lambda _{i}=(x_{i},X_{i},\psi )$, 
\begin{equation}
p(k|\mathcal{M},\mathcal{P},x_{i},X_{i},\psi )=p(k|\mathcal{M}%
,x_{i},X_{i},\psi )~,  \label{outcome prob d}
\end{equation}%
which states that, conditional on the information about $\mathcal{P}$ that
is codified into $\psi $, all \emph{other} details about $\mathcal{P}$ not
already conveyed by $\psi $ are irrelevant.

However, in a $\psi $-epistemic ontological model $\psi $ is not included in 
$\lambda $. The assumption there is that conditional on $\lambda _{i}$, for
any choice of $\mathcal{M}$, the outcome $k$ is \emph{completely independent
of all details about} $\mathcal{P}$ --- including any information that might
be codified into the epistemic $\psi $. \emph{It is these }$\psi $\emph{%
-epistemic ontological models that are shown by all the no-go theorems }\cite%
{Bell 1966}-\cite{Tumulka 2022}\emph{\ to disagree with QM}.

ED satisfies the causality assumption (\ref{prep prob}) but violates (\ref%
{outcome prob c}) and, therefore, \emph{ED is not an ontological model in
the sense of }\cite{Bell 1966}-\cite{Harrigan Spekkens 2010} which makes ED
immune to all the no-go theorems. More explicitly, the situation with ED is
the following: The ontic variables are positions of particles and/or
pointers, $\lambda =(x,X)$, and the information about the preparation
procedure $\mathcal{P}$ is fully conveyed by the wave function $\psi
(x_{i})\pi (X_{i})$. Then, the causality assumption, eq.(\ref{prep prob}),
reads 
\begin{equation}
p(\lambda _{i}|\mathcal{M},\mathcal{P})=p(x_{i},X_{i}|\psi )=|\psi
(x_{i})\pi (X_{i})|^{2}~,  \label{prep prob b}
\end{equation}%
where $\lambda _{i}=(x_{i},X_{i})$ are position values \emph{before}
entering the device $\mathcal{M}$. Next, since ED is a purely epistemic
dynamics, \emph{conditional on the epistemic wave function\ the distribution
of }$k$\emph{\ is independent of the ontic variables }$(x_{i},X_{i})$ ---
the latter have no causal influence on the future outcome $k$. Thus, instead
of (\ref{outcome prob c}), in ED the response function $p(k|\mathcal{M},%
\mathcal{P},\lambda _{i})$ is 
\begin{equation}
p(k|\mathcal{M},\mathcal{P},x_{i},X_{i})=p(k|\mathcal{M},\psi ,\pi )~.
\label{outcome prob e}
\end{equation}%
Substituting (\ref{prep prob b}) and (\ref{outcome prob e}) into (\ref%
{outcome prob b}) yields 
\begin{equation}
p(k|\mathcal{M},\mathcal{P})=p(k|\mathcal{M},\psi ,\pi )
\label{outcome prob f}
\end{equation}%
which agrees with the probability predicted by quantum mechanics. (See the
appendix for further details.)

It is worth emphasizing the difference between response functions in
ontological models, eq.(\ref{outcome prob c}), and in ED, eq.(\ref{outcome
prob e}): Ontological models assume that all the information about the
preparation that is relevant for the measurement is carried by the ontic
variables $\lambda _{i}$. This means that lurking in the background there is
an implicit assumption that there exists an ontic dynamics that relates the 
\emph{earlier} values of $\lambda _{i}$ as the system enters $\mathcal{M}$
to the \emph{later} values $\lambda _{f}$ that result in the outcome $k$.
The power of the no-go theorems lies in the fact that the details of the
ontic dynamics remain unspecified --- the ontic dynamics could be
deterministic or stochastic, it could be local or non-local, and so on ---
but some such dynamics must exist.

ED, on the other hand, makes a commitment to ontic states without making the
associated commitment to an ontic dynamics; the right hand side of (\ref%
{outcome prob e}) is fully epistemic. Wave functions do not guide the
particles; they only guide our expectations about where particles might be
found. \emph{ED is an epistemic} \emph{mechanics without an ontic mechanism}.

\noindent \textbf{Remark:} Beyond the general framework of ontological
models the various no-go theorems depend on additional assumptions. It is
these additional assumptions that have been offered as possible loopholes
(see \cite{Leifer 2014}). We emphasize that ED by-passes the issue of
additional assumptions and evades the ontological models framework at the
deeper level of the epistemic vs. ontic nature of the dynamics.

\noindent \textbf{Remark:} In \cite{Harrigan Spekkens 2010} $\psi $%
-epistemic models are defined as those that fail to be $\psi $-ontic. This
definition would in principle allow $\psi $-epistemic models based on exotic
probability theories \cite{Leifer 2014}. The ED approach shows that such
generalizations are not necessary; there is no need for exotic or even
quantum probabilities, or for a quantum logic, or for excess ontological
baggage, or for retrocausality. ED is $\psi $-epistemic in that $\psi (x)$
is directly related to the probabilities $\rho (x)$, that is, to knowledge
and beliefs, and to the conjugate momenta $\phi (x)$ that codify the
information that updates those probabilities.

\noindent \textbf{Remark:} The ideas above could have been formulated in a
more general setting of preparations that only determine a density matrix
and measurements that are described by positive operator valued measures
(POVMs). However, such increased generality would only serve to obscure the
main ideas and has no effect on the conclusions.

\section{Conclusion}

The solution of the problem of measurement within the ED framework hinges on
two points: first, entropic quantum dynamics is a theory of inference. The
issue of an unacceptable dichotomy of two modes of evolution --- continuous
unitary evolution versus discrete wave function collapse --- is resolved.
The two modes of evolution correspond to two modes of updating --- either by
a continuous maximization of an entropy or a discontinuous application of
Bayes' rule --- both of which, within the entropic inference framework, are
unified into a single entropic updating rule \cite{Caticha 2022}\cite%
{Caticha Giffin 2006}.

The second point is the privileged role of position --- particles (and also
pointer variables) have definite positions and therefore their preexisting
values are merely revealed and are not created by the act of measurement.
Other \textquotedblleft inferables\textquotedblright\ are introduced as a
matter of linguistic convenience to describe more complex experiments. These
inferables turn out to be attributes of the epistemic wave functions and not
of the ontic particles; their \textquotedblleft values\textquotedblright\
are indeed \textquotedblleft created\textquotedblright\ by the dynamical
process of measurement.

ED unscrambles Jaynes' proverbial omelette by imposing a clear-cut
separation between its ontic and epistemic elements. ED is a conservative
theory in that it attributes a definite ontic status to things such as
particles (or fields) and a definite epistemic status to probabilities and
wave functions; it is radically non-classical in that it denies the ontic
status of dynamics and of all observables except position.

\section*{Appendix}

Here we offer a more detailed analysis of eq.(\ref{outcome prob f}).

\noindent \textbf{The direct measurement of position}

The outcomes of the measurement are $k=x_{k}$, there is no pointer variable $%
X$ and the corresponding $\pi (X)$ can be omitted. The product of (\ref{prep
prob b}) and (\ref{outcome prob e}) is 
\begin{equation}
p(\lambda _{i}|\mathcal{M},\mathcal{P})p(k|\mathcal{M},\mathcal{P},\lambda
_{i})=|\psi (x_{i})|^{2}p(k|\mathcal{M},\psi )=|\psi (x_{i})|^{2}|\langle
s_{k}|\psi \rangle |^{2}
\end{equation}%
where we used (\ref{Born b}). Then (\ref{outcome prob b}) is 
\begin{equation}
p(k|\mathcal{M},\mathcal{P})=~\tint d^{3}x_{i}\,|\psi (x_{i})|^{2}|\langle
s_{k}|\psi \rangle |^{2}=|\langle s_{k}|\psi \rangle |^{2}~,  \label{Born e}
\end{equation}%
which agrees with the prediction of quantum mechanics.

\noindent \textbf{The indirect or von Neumann measurement}

The outcomes of the measurement are $k=(\alpha _{k},X_{f})$. The object of
interest is the eigenvalue $\alpha _{k}$, eq.(\ref{operator M}). The product
of (\ref{prep prob b}) and (\ref{outcome prob e}) is 
\begin{equation}
p(\lambda _{i}|\mathcal{M},\mathcal{P})p(k|\mathcal{M},\mathcal{P},\lambda
_{i})=|\psi (x_{i})\pi (X_{i})|^{2}p(\alpha _{k},X_{f}|\mathcal{M},\psi ,\pi
)
\end{equation}%
Using (\ref{Born b}) and (\ref{Born c}), the second factor on the right is 
\begin{equation}
p(\alpha _{k},X_{f}|\mathcal{M},\psi ,\pi )=|\langle s_{k},X_{f}|\hat{U}%
_{M}|\psi \pi \rangle |^{2}=|\langle s_{k}|\psi \rangle
|^{2}~N_{X}^{2}e^{-(X_{f}-\alpha _{k})^{2}/2\sigma _{\pi }^{2}}
\end{equation}%
substituting into (\ref{outcome prob b}) gives 
\begin{eqnarray}
p(\alpha _{k},X_{f}|\mathcal{M},\mathcal{P}) &=&~\tint
d^{3}x_{i}dX_{i}\,|\psi (x_{i})|^{2}|\pi (X_{i})|^{2}|\langle s_{k}|\psi
\rangle |^{2}N_{X}^{2}e^{-(X_{f}-\alpha _{k})^{2}/2\sigma _{\pi }^{2}} 
\notag \\
&=&|\langle s_{k}|\psi \rangle |^{2}N_{X}^{2}e^{-(X_{f}-\alpha
_{k})^{2}/2\sigma _{\pi }^{2}}~.
\end{eqnarray}%
Marginalizing over $X_{f}$ gives 
\begin{equation}
p(\alpha _{k}|\mathcal{M},\mathcal{P})=|\langle s_{k}|\psi \rangle |^{2}~,
\label{Born f}
\end{equation}%
which is the correct prediction according to quantum mechanics.

\noindent \textbf{Remark:} Equation (\ref{Born f}) gives the correct
probability but does not by itself describe the result of a measurement. The
latter consists of inferring the value $\alpha _{k}$ from the data $X_{f}$
that is actually observed. The relevant probability, $p(\alpha _{k}|\mathcal{%
M},\mathcal{P},X_{f})$, is given by Bayes theorem, 
\begin{equation}
p(\alpha _{k}|\mathcal{M},\mathcal{P},X_{f})=\frac{p(\alpha _{k},X_{f}|%
\mathcal{M},\mathcal{P})}{p(X_{f}|\mathcal{M},\mathcal{P})}~.  \tag{(35)}
\end{equation}%
Using (21) and (22) the result is 
\begin{equation}
p(\alpha _{k}|\mathcal{M},\mathcal{P},X_{f})=\frac{|\langle s_{k}|\psi
\rangle |^{2}N_{X}^{2}e^{-(X_{f}-\alpha _{k})^{2}/2\sigma _{\pi }^{2}}}{%
\sum\nolimits_{k^{\prime }}|\langle s_{k^{\prime }}|\psi \rangle
|^{2}N_{X}^{2}\,e^{-(X_{f}-\alpha _{k^{\prime }})^{2}/2\sigma _{\pi }^{2}}}~.
\tag{(36)}
\end{equation}%
When $\sigma _{\pi }$ is sufficiently small and the Gaussians are well
resolved eq.(22) tells us that with high probability we shall find values of 
$X_{f}$ concentrated at one of the discrete values $\alpha _{k}$. For $%
X_{f}\approx \alpha _{k}$ we find, 
\begin{equation}
p(\alpha _{k}|\mathcal{M},\mathcal{P},X_{f})\approx 1~,  \tag{(37)}
\end{equation}%
which means that a measurement of the pointer $X_{f}$ allows an immediate
inference of $\alpha $.

\subparagraph*{Acknowledgments}

I would like to thank D.T. Johnson, K. Vanslette, and S. Ipek for valuable
discussions and contributions at various stages of this program.


\begin{thebibliography}{99}
\bibitem{Bell 1990b} Bell, J. Against `measurement'.\ \emph{Physics World},
August \textbf{1990}, 33; reprinted in J. S. Bell \emph{Speakable and
Unspeakable in Quantum Mechanics} (Cambridge U. P., Cambridge, 2004).

\bibitem{Ballentine 1998} Ballentine, L. E. \emph{Quantum Mechanics: a
Modern Development }(World Scientific, 1998).

\bibitem{Schlosshauer 2004} Schl\"{o}sshauer, M. \textquotedblleft
Decoherence, the measurement problem, and interpretations of quantum
mechanics,\textquotedblright\ Rev. Mod. Phys. \textbf{76}, 1267 (2004).

\bibitem{Jaeger 2009} Jaeger, G. \emph{Entanglement, Information, and the
Interpretation of Quantum Mechanics} (Springer-Verlag, Berlin Heidelberg
2009).

\bibitem{Leifer 2014} Leifer, M. S. Is the Quantum State Real? An Extended
Review of $\Psi $-ontology Theorems.\ Quanta \textbf{2014}, \emph{3}, 67;
arXiv.org:1409.1570.

\bibitem{Caticha 2019} Caticha, A. The Entropic Dynamics approach to Quantum
Mechanics. \emph{Entropy }\textbf{2019}, \emph{21}, 943;
arXiv.org:1908.04693.

\bibitem{Caticha 2021} Caticha, A. Quantum mechanics as Hamilton-Killing
flows on a statistical manifold. Phys. Sci. Forum \textbf{2021}, 3, 12;
arXiv:2107.08502.

\bibitem{Caticha 2022} Caticha, A. \emph{Entropic Physics: Probability,
Entropy, and the Foundations of Physics}; available online at \emph{%
https://www.arielcaticha.com/} (accessed on June 30, 2022).

\bibitem{Johnson Caticha 2011} Johnson, D.T., Caticha, A. Entropic dynamics
and the quantum measurement problem.\ AIP Conf. Proc. \textbf{2012}, 1443,
104; arXiv:1108.2550.

\bibitem{Vanslette Caticha 2016} Vanslette, K., Caticha, A. Quantum
measurement and weak values in entropic quantum dynamics.\ AIP Conf. Proc. 
\textbf{1853}, 090003 (2017); arXiv:1701.00781.

\bibitem{Caticha 2000} Caticha, A. Insufficient reason and entropy in
quantum theory. \emph{Found. Phys. \textbf{2000}},\emph{\ }30, 227
arXiv.org/abs/quant-ph/9810074.

\bibitem{Bell 1966} Bell, J. S. On the Problem of Hidden Variables in
Quantum Mechanics. \emph{Rev. Mod. Phys.} \textbf{1966}, 38(3), 447-452.

\bibitem{Spekkens 2005} Spekkens, R. W. Contextuality for preparations,
transformations and unsharp measurements. \emph{Phys. Rev. A \textbf{2005}},%
\emph{\ 71}, 052108.

\bibitem{Harrigan Spekkens 2010} Harrigan, N., Spekkens, R. W. Einstein,
Incompleteness, and the Epistemic View of Quantum States. \emph{Found. Phys. 
\textbf{2010}},\emph{\ 4}0, 125-157.

\bibitem{Bell 1990a} Bell, J. S., La nouvelle cuisine. Reprinted in J. S.
Bell \emph{Speakable and Unspeakable in Quantum Mechanics} (Cambridge U. P.,
Cambridge, 2004).

\bibitem{Hardy 2004} Hardy, L., Quantum ontological excess baggage. \emph{%
Stud. Hist. Phil. Mod. Phys.} \textbf{2004}, 35, 267-276.

\bibitem{Montina 2008} Montina, A., Exponential complexity and ontological
theories of quantum mechanics. \emph{Phys. Rev. A \textbf{2008}},\emph{\ 77}%
, 022104.

\bibitem{Pusey et al 2012} Pusey, M. F., Barrett, J., Rudolph, T. On the
reality of the quantum state.\ \emph{Nature Physics} \textbf{2012},\ 8,
475-478; arXiv:1111.3328.

\bibitem{Colbeck Renner 2012} Colbeck, R., Renner, R., Is a System's Wave
Function in One-to-One Correspondence with Its Elements of Reality? \emph{%
Phys. Rev. Lett.} \textbf{2012}, 108, 150402.

\bibitem{Hardy 2012} Hardy, L., Are quantum states real? \emph{Int. J. mod.
Phys. B} \textbf{2013}, 27, 1345012; arXiv:1205.1439.

\bibitem{Schlosshauer Fine 2014} Schlosshauer, M,. Fine, A., $No-go$ theorem
for the composition of quantum states. \emph{Phys. Rev. Lett.} \textbf{2014}%
, 112, 070407.

\bibitem{Leifer 2014b} Leifer, M., $\psi $-epistemic models are
exponentially bad at explaining the distinguishability of quantum states. 
\emph{Phys. Rev. Lett.} \textbf{2014}, 112, 160404.

\bibitem{Barrett et al 2014} Barrett, J., Cavalcanti, E.G., Lal, R.,
Maroney, O., No $\psi $-epistemic model can fully explain the
indistinguishability of quantum states. \emph{Phys. Rev. Lett.} \textbf{2014}%
, 112, 250403.

\bibitem{Branciard 2014} Branciard, C., How $\psi $-epistemic models fail at
explaining the indistinguishability of quantum states. \emph{Phys. Rev. Lett.%
} \textbf{2014}, 113, 020409.

\bibitem{Ruebeck et al 2020} Ruebeck, J., Lillystone, P., Emerson, J., $\psi 
$-epistemic interpretations of quantum theory have a measurement problem.
Quantum \textbf{2020}, 4, 242; arXiv:1812.08218.

\bibitem{Tumulka 2022} Tumulka, R. Limitations to Genuine Measurements in
Ontological Models of Quantum Mechanics; arXiv:2205.05520.

\bibitem{Demme Caticha 2016} Demme, A., Caticha, A., The Classical Limit of
Entropic Quantum Dynamics.\ AIP Conf. Proc. \textbf{2017}, 1853, 090001;
arXiv.org:1612.01905.

\bibitem{Caticha Giffin 2006} A. Caticha and A. Giffin, \textquotedblleft
Updating Probabilities\textquotedblright ,\ \emph{Bayesian Inference and
Maximum Entropy Methods in Science and Engineering}, ed. by A.
Mohammad-Djafari, AIP Conf. Proc. \textbf{872}, 31 (2006); arXiv.org/
abs/physics/0608185.
\end{thebibliography}
\end{document}